\documentclass[twocolumn,prl,aps]{revtex4}
\usepackage{amsmath}
\usepackage{amssymb}
\usepackage{graphicx}
\usepackage{dcolumn}
\usepackage{graphicx}
\usepackage{dcolumn}
\usepackage{bm}

\begin{document}

\preprint{Phys.Rev.Lett.}

\title{Microwave induced nonlocal transport in two-dimensional electron system.}
\author{A. D. Levin,$^1$ Z. S. Momtaz, $^1$ G. M. Gusev,$^1$
 and A. K. Bakarov, $^{2}$}

\affiliation{$^1$Instituto de F\'{\i}sica da Universidade de S\~ao
Paulo, 135960-170, S\~ao Paulo, SP, Brazil}
\affiliation{$^2$Institute of Semiconductor Physics, Novosibirsk
630090, Russia}

\date{\today}
\begin{abstract}
 We observe microwave induced nonlocal resistance in magnetotransport
 in single and bilayer electronic systems. The obtained results
 provide evidence for an edge state current stabilized by
 microwave irradiation due to nonlinear resonances.
 Our observation are closely related to
 microwave induced oscillations and zero resistance states in a
 two-dimensional (2D) electron system.

\end{abstract}

\maketitle

A few years ago, a new class of the non-equilibrium phenomena was
observed, when an ultrahigh mobility 2D electron gas, subjected to a
weak magnetic field was also irradiated with microwaves
\cite{dmitriev}. This included microwave-induced resistance
oscillations (MIROs) \cite{zudov} and a zero-resistance state
(ZRS)\cite{mani}. Many microscopic mechanisms of MIRO have been
proposed, mainly originating from the scattering-assisted electron
transitions between different Landau levels in the presence of
microwave excitation. The most developed theories account for
spatial displacement of electrons along the applied dc field under
scattering-assisted microwave absorption ("displacement" mechanism)
\cite{ryzhii,durst}, and an oscillatory contribution to the
isotropic part of the electron distribution function ("inelastic"
mechanism) \cite{dmitriev1}. Both these mechanisms describe the
periodicity and phase of MIROs observed in experiments and can lead
to an absolute negative conductivity $ \sigma < 0$.  ZRS emerges
from the instability of a homogeneous state with  $ \sigma < 0$ and
the nonequilibrium phase transition into a domain state with zero
net resistance \cite{andreev}. Two more alternative approaches to
the MW-induced effects in dissipative resistance, such as the
radiation-driven electron-orbit \cite{inarrea} and near-contact
region \cite {mikhailov} models, have been recently proposed.

A striking similarity has been emphasized between QHE and ZRS: both
effects exhibit vanishing longitudinal resistance $R_{xx}$, when the
propagation along the sample edge is ballistic, although the
magnetic field intensity is quite different. One naturally expects
that a strong magnetic field stabilizes edge states, and, therefore,
that the QHE is robust against disorder \cite{buttiker}. It has been
shown that microwave radiation can also stabilize guiding along
sample edges in the presence of a relatively weak magnetic field
leading to a ballistic dissipation-less transport regime, which also
results in vanishing $R_{xx}$ \cite{chepelyanskii}. Indeed such
transport is much less robust than those in the QHE regime and
requires samples with ultrahigh electron mobility. This model
also avoids the fundamental assumption made in those approaches
\cite{ryzhii,durst, dmitriev1} that cyclotron harmonic absorption at
high $j$ can be explained by the presence of the short range
potential, while in high mobility samples the long range potential
plays a dominant role.

The method used for probing the property of the edge states is nonlocal
electrical measurement. If a finite voltage is applied between a pair of the probes, a
net current appears along the sample edge, which can be detected by
another pair of voltage probes far away from the bulk current
path.

In this letter we present studies of the nonlocal resistance in
narrow  (NQW)  and  wide  (WQW) quantum wells, which represent
single and two-subband 2D systems respectively, exposed to microwave
irradiation. We find a relatively large  ($ \sim 0.05\times R_{xx}$)
nonlocal resistance in the vicinity of the ratio $j \approx
\omega/\omega_{c} \approx 3.15/4$, where $\omega$ is the radiation
angular frequency, $\omega_{c}=eB/m$ is the cyclotron angular
frequency, and $m$ is the effective mass of the electrons. We
attribute the observed nonlocality to the existence of edge states
stabilized by microwave irradiation and a weak magnetic field. We
provide a model taking into account the edge and bulk contributions
to the total current in the local and nonlocal geometries.
\begin{figure}[ht!]
\includegraphics[width=9cm,clip=]{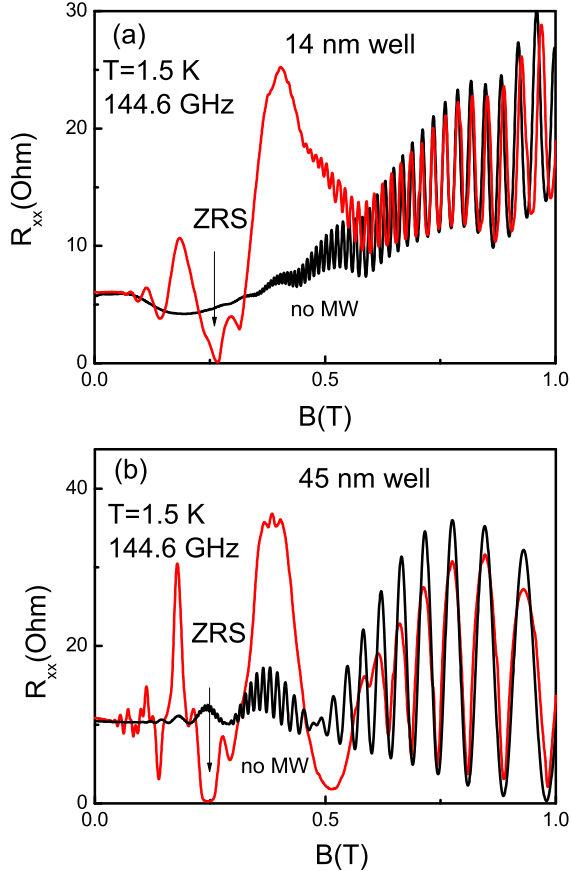}
\caption{\label{fig.1}(Color online) Longitudinal $R_{xx}$ ( I=1,4;
V=2,3) resistance without (no MW) and with  microwave irradiation
(144,6 GHz) in  (a) a narrow (14 nm) and a (b)  wide (45 nm) quantum
well. Arrows indicate the regions of vanishing resistance.}
\end{figure}

We have studied both narrow (14~nm) and wide (45~nm) quantum wells
with an electron density of $n_{s} \simeq 1.0 \times 10^{12}$
cm$^{-2}$ and a mobility of $\mu~\simeq 1.7-3.2 \times 10^{6}$
cm$^{2}$/V s, respectively, at temperature of 1.4 K and after a brief illumination
with a red diode. Owing to charge redistribution, WQWs with high
electron density form a bilayer configuration, i.e. two wells near
the interfaces are separated by an electrostatic potential barrier
and two subbands appear as a result of tunnel hybridization of 2D
electron states (symmetric  and antisymmetric), which are separated
 in energy by $\Delta_{SAS}$. In NQW electrons also occupy
two subbands after illumination, but the carrier density of the
second subband is much smaller than the density of the lower
subband. We have measured resistance on two different types of the
devices. Device A is a conventional Hall bar patterned structure
(length $l~\times$ width $w$=500~$\mu$m$\times$200~$\mu$m) with six
contacts for identifying nonlocal transport over macroscopic
distances. Device B is designed for multi-terminal measurements. The
sample consists of three $5 \mu m$ wide consecutive segments of
different length ($5, 15 , 5 \mu m$), and 8 voltage probes
\cite{supplementary}. The measurements have been carried out in a
VTI cryostat with a waveguide to deliver MW irradiation (frequency
range 110 to 170 GHz) down to the sample and by using a conventional
lock-in technique to measure the longitudinal resistance $R_{xx}$.
In WQW the value of $\Delta_{SAS}$~=~1.40 meV is extracted from the
periodicity of low-field MIS oscillations, see Ref. \cite{mamani}.
Several devices from the same wafers have been studied.

\begin{figure}[ht!]
\includegraphics[width=9cm,clip=]{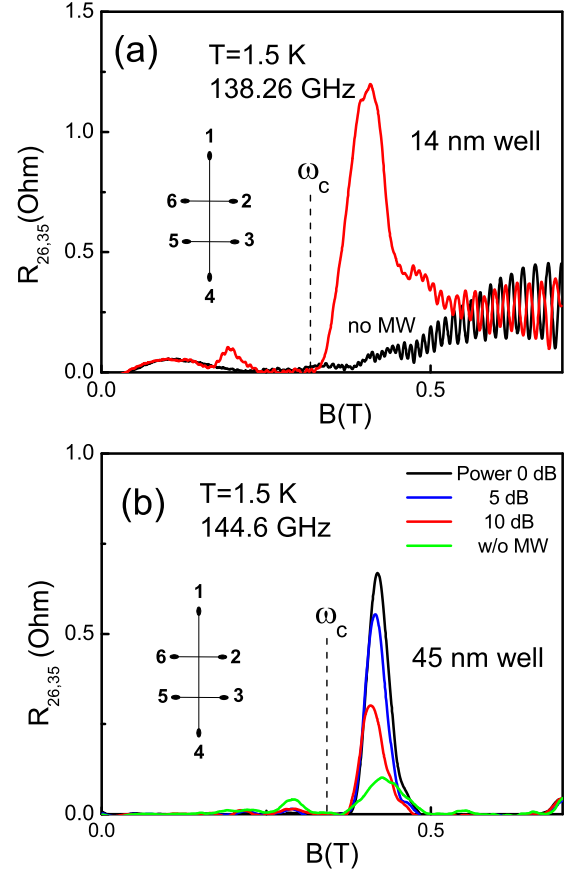}
\caption{\label{fig.2}(Color online) (a) Nonlocal  resistance
$R_{26,35}$ (I=2,6; V=3,5) under microwave irradiation (143 GHz)in a
wide (45 nm) quantum well with decreasing microwave power. (b)
Nonlocal resistance $R_{26,35}$ (I=2,6; V=3,5) without (no MW) and
under microwave irradiation (144,6 GHz) in a  narrow (14 nm) quantum
well. Insets show the measurement configuration.}
\end{figure}

In Fig. \ref{fig.1} we present dark resistance and the observation
of a ZRS (marked with an arrow) for 144,6 GHz and at a temperature
of 1.5 K in single layer and bilayer Hall bar devices. In the
presence of the microwave irradiations MIRO appear in NQW, and one
of the minimums develops into ZRS. The resistance of both quantum
wells reveals magnetointesubband (MIS) oscillations caused by the
periodic modulation of the probability of intersubband transitions
by the magnetic field (see Ref. \cite{mamani}). Note, however, that
MIS oscillations in the narrow well are observed at a relatively
high magnetic field due to the low second subband density, and are
therefore, they are almost unaffected by MW radiation. In contrast,
exposing a bilayer system to MW leads to interference of the MIRO
and MIS oscillations with enhancement, suppression or inversion of
the MIS peak correlated with MW frequency \cite{wiedmann1}.
Moreover, a zero resistance state develops from the MIS maximum
\cite{wiedmann2}.

\begin{figure}[ht!]
\includegraphics[width=9cm,clip=]{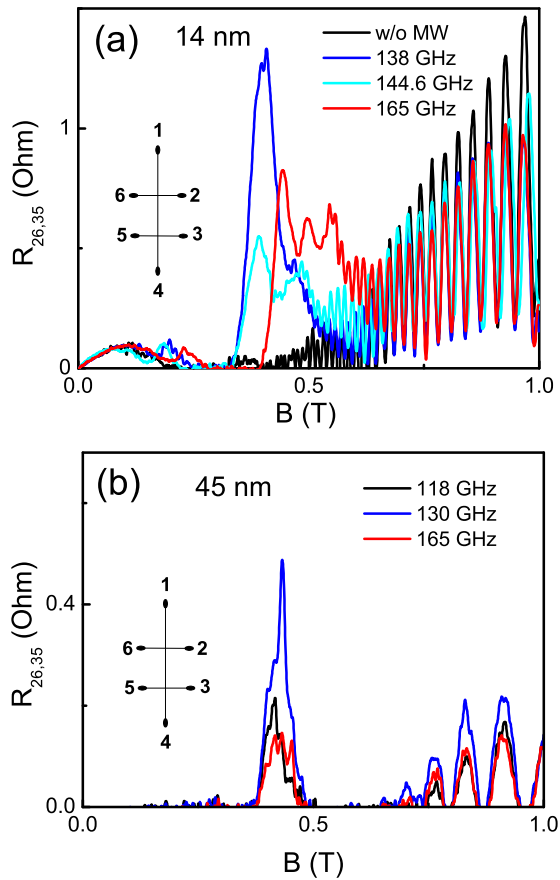}
\caption{\label{fig.3}(Color online)  Nonlocal  $R_{26,35}$ (I=2,6; V=3,5)
resistances for a narrow (a) and wide (b) quantum wells and   for different  microwave frequencies.
 Insets show the measurement configuration.}
\end{figure}

For the same samples with a Hall bar pattern, the nonlocal
resistance $R_{NL}=R_{26,35}$ (I=2,6; V=3,5) was also measured.
Figure 2 shows $R_{26,35}$ for both types of quantum wells (WQW and
NQW) in the presence of microwave irradiation and at different
intensities of radiation. Both samples display a prominent peak in
nonlocal resistance corresponding to a peak in $R_{xx}$ around $j
\approx 3.15/4$. However, examination of figures 1 and 2 reveals a
drastic difference between local and nonlocal effects. In
particular, the second peak at $B\approx 0.18 T$ in local
resistance, which has almost the same amplitude as the peak near 0.4
T, vanishes in the nonlocal resistance for WQW. Figure 3 illustrates
microwave -induced nonlocal resistance for three chosen frequencies.
One can see only one dominant peak near $B\approx 0.4 T$. The
magnitude of the peak varies with frequency due to the variation in
microwave power. The position of the peak in NQW is correlated with
frequency, while, in the bilayer system, peaks developed from
combined MIS-MIR oscillations, and therefore, their location depends
on subband splitting and is less sensitive to frequency
\cite{wiedmann2}.The classical ohmic contribution to the nonlocal
effect is given by $R^{classical}_{NL}\approx\ \rho_{xx}\exp(-\pi
L/w)$ for narrow strip geometry where $L$ is the distance between
the voltage probes, and $w$ is the strip width \cite{pauw}. For our
geometry we estimate $R^{classical}_{NL}/R_{xx}\approx
3\times10^{-4}$ for a zero magnetic field. In the QHE regime, the
nonlocal resistance $R_{NL}$ arises from the suppression of electron
scattering between the outermost edge channels and the
backscattering of the innermost channel via the bulk states
\cite{buttiker, mcEuen, dolgopolov}. It appears only when the
topmost Landau level is partially occupied, and scattering via bulk
states is allowed.

We have also measured the nonlocal response in device B in other
geometries and found similar behavior. Note, however, that these
samples have less mobility ($~0.9\times10^{6} cm^{2}/Vs$), and
demonstrate microwave-induced nonequilibrium oscillations of smaller
amplitude, without reaching the zero resistance state. These and
other measurements described in \cite{supplementary} provide
evidence for microwave induced edge state transport in the low
magnetic field regime. Nonzero nonlocal resistance implies that the
dissipation-less edge state  transport persists over macroscopic
distances, because the length of the edge channels are determined by
the distance between the metallic contacts ($\sim 1 mm$), or at
least, by the distance between potential probes ($\sim 0.5 mm$).

\begin{figure}[ht!]
\includegraphics[width=9cm,clip=]{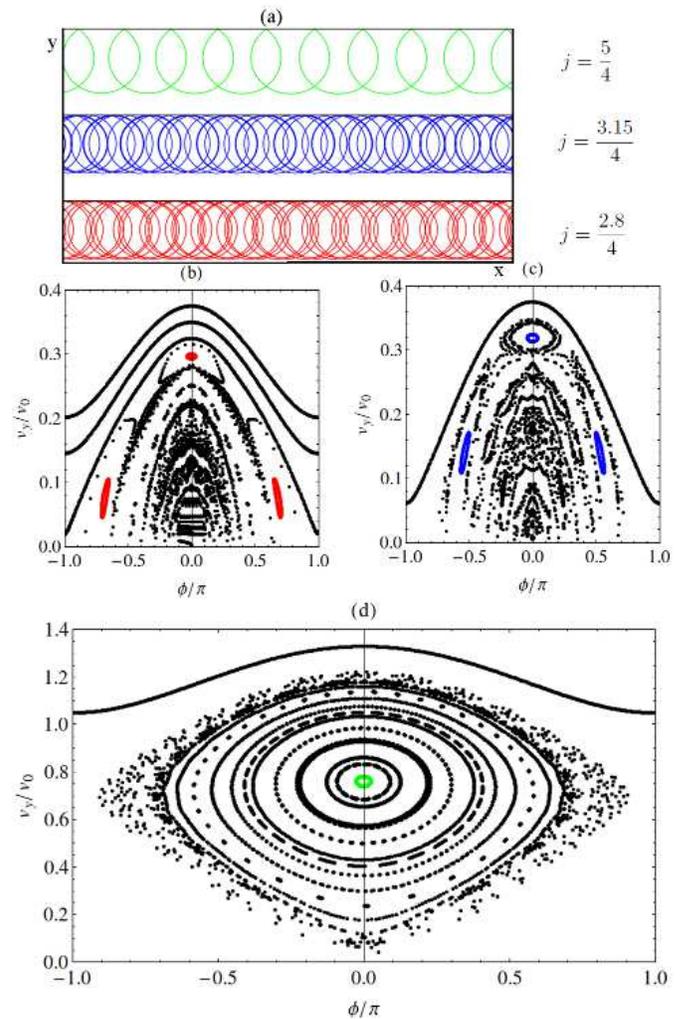}
\caption{\label{fig.4}(Color online)  (a) Examples of electron
trajectories along the sample edge for several values of $j=5/4$,
$3.15/4$, $2.8/4$; (b) Poincar$\acute{e}$ section of for $j=2.8/4$,
(c) $j=3.15/4$, (d) $j=5/4$ at y-polarized field with $\varepsilon
=0.02$.}
\end{figure}
The nonlocal effect described above can be understood within a
common framework based on modern nonlinear dynamics. As was
indicated in \cite{chepelyanskii} for propagating edge channels, a
microwave field creates a nonlinear resonance well described by the
standard map, known as the Chirikov standard map \cite{chirikov}. In
our case it is constructed by a Poincar$\acute{e}$´s surface of a
section for electrons moving in the vicinity of the sample edge
modeled as a specular wall in the presence of the microwave driving
field. The details of the model are described in
\cite{chepelyanskii}.  In order to compare the theory with our
experiment, we extend the results of this model to our specific
sample parameters and experimental conditions \cite{supplementary}.
We are mostly interested in the dynamics of electrons in the
vicinity of the ratio $j=3.15/4$, where the microwave induced peak
in the nonlocal response is observed (figs. 2,3). Figure 4a shows
electron trajectories in the edge vicinity for different values of
the ratio $j$. One can see that the microwave field strongly
modifies the dynamics along the edge. The figures 4 b,c,d show
Poincar$\acute{e}$ sections for the wall model and different values
of the magnetic field corresponding to the peak in $R_{NL}$ around
$B=0.42 T$ ($j=3.15/4$), on the high- field side of the peak
($j=2.8/4$) and on the resistance minima ($j=5/4$). For $j=5/4$
Poincar$\acute{e}$ sections exhibit periodic and quasiperiodic
trajectories surrounded by a chaotic sea. For $j=2.8/4$ and
$j=3.15/4$ Poincar$\acute{e}$ sections exhibits less stable
dynamics, but a periodic component remains present. The existence of
the periodic orbits plays a fundamental role in the local and
nonlocal resistivity of a 2D gas. The truly dissipation-less edge
channels may carry the same electrochemical potential  $\mu$ over a
macroscopic distance to the different voltage probes. As a
consequence, $\Delta \mu =0$ which results in vanishing $R_{xx}$ and
$R_{NL}$. If a certain fraction of the edge states are scattered
into the bulk, it leads to different local chemical potentials,
finite $\Delta \mu$ and resistivity. This situation closely
resembles the QHE, when it is possible to treat the edge and bulk
conducting pathways separately. This may lead to nonlocal
resistivity. For example, in the QHE regime the nonlocal resistance
$R_{NL}$ arises from the suppression of electron scattering between
the outermost edge channels and backscattering of the innermost
channel via the bulk states \cite{buttiker, mcEuen, dolgopolov}.
It is worth noting that the actual shape of the wall potential is rather
 parabolic than the hard wall. We have compared nonlinear resonance and Poincar$\acute{e}$ sections
 for both potential and found no difference  in the dynamics of electrons (for details see \cite{supplementary}).

However, demonstration of the existence of the periodic orbits
stabilized by the MW field is not enough to justify the nonlocal
response and some further qualitative analysis might be required to
compare the magnitude of $R_{NL}$ with calculations using the simple
model. In the rest of the paper, we provide a model describing the
edge and bulk contribution to local and nonlocal resistance
\cite{supplementary}. The mechanism responsible for the observed
peak in local and nonlocal magnetoresistance near $j=3.15/4$ relies
on the combination of the edge state and bulk transport
contributions, when bulk-edge coupling  is taken into account
\cite{dolgopolov2}. The bulk electrons and edge modes are
consequently described by the local chemical potentials $\psi$ and
$\varphi$. We can introduce the phenomenological constant $g$, which
represent scattering between modes $\varphi$ and $\psi$. The edge
state transport can be described by equations for particle density
\cite{dolgopolov, dolgopolov2}, taking into account the scattering
between edge modes and the bulk \cite{supplementary}. The model
reproduces the experimental values of the  local $R_{14,23}\approx
40$ ohm and nonlocal $R_{26,35}\approx 1.2$ ohm resistances with
adjustable parameter $g=0.005 \mu m^{-1}$. Note, however, that more
precise calculations require exact knowledge of the fractions of
electrons channeling along the wall $P$. Taking into account the
total number of the Landau levels near $B\approx0.42$ T,  $N\approx
120$, we may choose $M=P\times N\approx 1-3$ and a current carried
by edge channels $I\sim Me^{2}\varphi /h $.

A consensus, based on many experimental studies, including magnetic
field, frequency, temperature etc dependencies, now exists
concerning the "inelastic"  mechanism, which may by now be
considered as the dominant contribution to MIRO and ZRS in a high
mobility electron system \cite{dmitriev}. Moreover, observation
of ZRS in Corbino geometry \cite{yang} and contactless \cite{andreev} measurements strongly indicates that
microwave induced nonequilibrium effect is the bulk rather than the edge state phenomenon.
Our finding may  indicate that MIRO and ZRS are a
very rich physical phenomenon, which results from a combination of the
both bulk and edge state contributions. We believe that ZRS phenomena is
somewhat like a quantum Hall effect (although not exactly the same),
 which can be described as a bulk or/and edge phenomena
 (see for example \cite{kao}). Indeed both descriptions are experimentally
 supported  from measurements: the observation  of the nonlocal effects clearly
  demonstrates edge-state conduction \cite{mcEuen},  and observation of the
  charge transfer in Corbino geometry, where  edge transport is shunted via
  concentric contacts, show that the quantum Hall effect as a consequence of pure bulk transport
  is possible \cite{dolgopolov3}.

From our experiment we may conclude that the edge state effect is dominant or comparable
with bulk contribution near $\omega/\omega_{c} \approx 3.15/4$. Note, however, that 
our result does not explicitly rule out the bulk mechanism  near minimum $j=5/4$ 
and, therefore, does not contradict with previous investigations.

In conclusion, we have observed a microwave induced nonlocal
magnetoresistance peak in the vicinity of the ratio
$\omega/\omega_{c} \approx 3.15/4$. This data offer evidence that,
in a low magnetic field, MW induced edge-state transport really
extends over a macroscopic distance of $\sim 1 mm$. We compare our
results to a transport model that takes into account the combination
of the edge state and the bulk transport contributions and the
backscattering within the bulk-edge coupling.

 We thank Z.D.Kvon for helpful
discussions. The financial support of this work by FAPESP, CNPq
(Brazilian agencies) is acknowledged.

\end{document}